# Coexistence of Logarithmic and SdH Quantum Oscillations in Ferromagnetic Cr-doped Tellurium Single Crystals

Shu-Juan Zhang[1,*], Lei Chen[2], Shuang-Shuang Li[3], Ying Zhang[3], Jian-Min Yan[4], Fang Tang[5], Yong Fang[5], Lin-Feng Fei[3], Weiyao Zhao[6], Julie Karel[6], Yang Chai[4], and Ren-Kui Zheng[2,*]

[1]School of Materials and Mechanic & Electrical Engineering, Jiangxi Science and Technology Normal University, Nanchang 330038, China

[2]School of Physics and Materials Science, Guangzhou University, Guangzhou 510006, China

[3]School of Materials Science and Engineering and Jiangxi Engineering Laboratory for Advanced Functional Thin Films, Nanchang University, Nanchang 330031, China

[4]Department of Applied Physics, The Hong Kong Polytechnic University, Hong Kong 999077, China

[5]Jiangsu Laboratory of Advanced Functional Materials, Department of Physics, Changshu Institute of Technology, Changshu 215500, China

[6]Department of Materials Science & Engineering, & ARC Centre of Excellence in Future Low-Energy Electronics Technologies, Monash University, Clayton VIC 3800, Australia

We report the synthesis of transition-metal-doped ferromagnetic elemental single-crystal semiconductors with quantum oscillations using the physical vapor transport method. The 7.7 atom% Cr-doped Te crystals (Cr:Te) show ferromagnetism, butterfly-like negative magnetoresistance in the low temperature (< 3.8 K) and low field (< 0.15 T) region, and high Hall mobility, e.g., 1320 cm$^2$ V$^{-1}$ s$^{-1}$ at 30 K and 350 cm$^2$ V$^{-1}$ s$^{-1}$ at 300 K, implying that Cr:Te crystals are ferromagnetic elemental semiconductors. When $B$ // [001] // I, the maximum negative MR is ∼ -27% at $T$ = 20 K and $B$ = 8 T. In the low temperature semiconducting region, Cr:Te crystals show strong discrete scale invariance dominated logarithmic quantum oscillations when the direction of the magnetic field $B$ is parallel to the [100] crystallographic direction ($B$ // [100]) and show Landau quantization dominated Shubnikov-de Haas (SdH)

---

E-mail: *zrk@ustc.edu; *shujuan866@163.com

oscillations for *B* // [210] direction, which suggests the broken rotation symmetry of the Fermi pockets in the Cr:Te crystals. The findings of coexistence of multiple quantum oscillations and ferromagnetism in such an elemental quantum material may inspire more study of narrow bandgap semiconductors with ferromagnetism and quantum phenomena.

**Introduction**

Quantum oscillations are periodic modulation of certain physical parameters by magnetic fields, which is a manifestation of the electronic band structures, therefore drawing increasing attention in condensed matter physics. The first reported quantum oscillation is related to the Landau quantization of the magnetization of metals in magnetic fields, predicated by Landau. Experimentally, Shubnikov and de Haas observed the oscillation of the magnetoresistance in bismuth single crystals,[1] independent from Landau's prediction, namely, the Shubnikov-de Haas (SdH) oscillation. The periodic oscillations due to the Landau quantization effect were also observed in the magnetic field dependent magnetization,[2] specific heat,[3] and Seebeck effect[4] measurements. The periodicity of SdH oscillations is usually plotted against inverse magnetic field 1/*B*, which means that the oscillations are evenly spaced on a 1/*B* scale.[5] In mesoscopic ring or cylinder structures, the quasiparticle quantum interference gives rise to another type of oscillation in magnetotransport behaviors, namely, the Aharonov-Bohm effect,[6,7] in which the oscillation is evenly spaced with *B*. Recently, another quantum oscillation with its periodicity on a log*B* scale was reported in ZrTe$_5$[8] and HfTe$_5$[9] single crystals beyond their quantum limit. The log-periodic quantum oscillations are the reminiscent of the discrete scale invariance (DSI) behavior, which indicates that the system possesses a geometric series of length scales. In a system with charge carriers from both trivial bands and Dirac bands, the long-range Coulomb attraction follows $R^{-1}$ rule, where $R$ denotes the distance of Coulomb interaction. Therefore, the massless Dirac Hamiltonian with the Coulomb interaction obeys the scale invariance.[8] The quasi-bound states in such a system induces resonant scattering process in the transport property at the Fermi level, which leads to a log-periodic correction to the MR.[8]

In the last few years, the elemental tellurium semiconductor with a nontrivial band structure, excellent device performance and multi-type quantum oscillations has drawn increasing attention.[10-20] Tellurium crystallizes in the trigonal structure with $P3_121$ or $P3_221$ space group, depending on the left-/right-handed screw axis. Theoretically, the calculated conduction bands of tellurium have a spin splitting similar to the Rashba splitting around the $H$ point in the reciprocal space, but unlike the Rashba splitting, the spin texture of which likes a hedgehog.[2,20] Experimentally, Akiba *et al.* realized the Weyl metallic phase by applying external pressure to tellurium.[12] Qiu *et al.* observed the quantum Hall effect and nontrivial Berry phase in *n*-type semiconducting tellurene.[3] Moreover, tellurium-based field-effect transistors have been fabricated to demonstrate its potential applications in electronics.[4] The magnetotransport properties of tellurium single crystals has also been studied in magnetic fields up to 55 T and pressures up to 5 GPa, which reveals pressure-induced topological transition from a semiconductor to a Weyl semimetal[5,6] and the existence of DSI-induced logarithmic quantum oscillations above the quantum limit fields. To explore more magnetotransport behavior in this system, we further doped transition metal into the Te single crystals.

Here, we report magnetotransport and magnetic properties of high-quality transition-metal Cr doped tellurium $Cr_xTe_{1-x}$ (Cr:Te) single crystals grown by the physical vapor transport method. Depending on the direction of the magnetic field with respect to the crystallographic directions of Te crystals, quantum oscillations following different oscillation models were observed, i.e., the quantum oscillation is periodic with log$B$ for $B$ // [100] direction and is periodic with 1/$B$ for $B$ // [210] direction. With Cr doping, the Cr:Te single crystals show ferromagnetism and butterfly-like negative magnetoresistance (MR) in low temperatures. The Cr:Te crystals show comparable high Hall mobility with undoped tellurium, which supports the feasibility of magnetic doping technique in Te-based semiconductors.

**Experiments**

The crystal growth process is detailed here: 4N purity Cr and Te powders with atomic ratio of 7:93 was sealed in a silica tube in vacuum ($2\times10^{-3}$ Pa). The sealed mixture was placed in a two-temperature-zone furnace, followed by: 1) heating to 1000 °C at 1 °C/min, 2) keeping at 1000 °C for 1 hour, 3) cooling down to 400 °C and 300 °C. respectively, 4) keeping the 100 °C temperature gradient for 1 week. After these growth process, needle-like Cr:Te single crystals with a size of ~ 5-10×1×1 mm$^3$ were obtained.

X-ray diffraction (XRD) patterns were collected using a Rigaku SmartLab X-ray diffractometer equipped with Cu K$\alpha_1$ radiation ($\lambda$=1.5406 Å). The surface and cross-sectional microstructure, chemical composition, and elemental mapping were characterized via the energy dispersive spectroscopy (EDS) and scanning electron microscopy (SEM) using a FEI focused ion beam system (Thermo Fisher Scientific, Scios2) equipped with an EDS module (Oxford AztecLive UltimMax 80). High-resolution transmission electron microscopy (HRTEM) and selected area electron diffraction (SAED) pattern were obtained using a JEOL JEM-F200 transmission electron microscope.

The magnetotransport properties were measured using a physical property measurement system (PPMS) (DynaCool-14, Quantum Design). Six ohmic contacts in a Hall-bar configuration were fabricated on the shiny surface of single crystals using indium. For magnetotransport measurements, the electric current is always along the length direction (i.e., the $c$ axis) of crystals and the direction of the magnetic field is applied along the [100] (i.e., the $a$ axis), [210], and [001] (i.e., the $c$ axis) crystallographic directions, respectively. The magnetic properties were measured using a superconducting quantum interference device (SQUID) (MPMS3, Quantum Design), with the direction of the magnetic field along the [100] direction.

**Results and Discussion**

Figure 1(a) shows the crystal structure of the tellurium with helical atomic chains. The quasi-1D chains were bonded by the van de Waals force, in which the tellurium atoms propagate along the $c$ axis. The top view of the crystal structure is shown in Fig. 1(b), where the crystallographic [100], [210], and [010] directions, together with the $a$, $b$, and $c$ axes, are indicated by red arrows. A photograph of an as-grown Te single crystal with a length of ~6 mm is shown in the inset of Fig. 1(c). Its cross-sectional SEM image is shown in Fig. 1(d), which demonstrates the inequilateral hexagon shape with rough 120° rotation symmetry. As the pink arrows indicated, the (100) and (0̄10) planes possess different cleavage surface size, indicating different free energy of atom stacking. The XRD θ-2θ scan measurements were performed on the (100) plane of a crystal. As shown in Fig. 1(c), sharp comb-like ($l$00) diffraction peaks are observed. A rocking curve scan taken around the (100) diffraction peak yields a full width at half maximum (FWHM) of 0.03° [Fig. 1(e)], indicating high crystalline quality of the crystal. The elemental mapping via EDS were performed on the (100) plane (Supplementary Fig. S1) and the cross-sectional area [the white box (12×8 μm) in Fig. 1(f)] of as-grown single crystals, respectively. As shown in Figs. 1(g) and 1(h), the Cr dopant is uniformly distributed within the cross-sectional area of the crystal, with an overall Cr concentration of ~7.7 atom% [Fig. 1(i)], which is 10% larger than the nominal doping level of 7 atom%. The slight off-stoichiometric effect was also reported in other single crystals that were grown via vapor-transport-based single-crystal growth methods.[21,22] Fig. 1(j) shows a typical atomically resolved HRTEM image of a Cr:Te crystal. The interplanar spacings are 5.65 Å and 2.15 Å, corresponding to the Te (001) and (110) planes, respectively. The in-plane $d$ spacing from HRTEM results is slightly smaller than those of Te single crystals (2.22 Å),[20] which is probably due to the replacement of a portion of Te atoms by Cr atoms. Both the HRTEM image and the SAED [Fig. 1(k)] confirm the high crystalline quality of the Cr:Te crystals.

Magnetotransport measurements were performed on such a rod-like single crystal. As shown in Fig. 2(a), the temperature dependent resistivity (RT) was studied in the temperature region from 1.8 to 300 K, with the direction of the magnetic field $B$ along the [210] direction of the crystal. For $B$=0 T, the resistivity shows a minimum at $T_{\min} \sim 39$ K, above which the crystal behaves like a metal with the resistivity between 10 and 30 mΩ cm. Below 39 K, the resistivity increases rapidly with cooling and follows the Arrhenius equation $\ln\rho = \ln\rho_0 + E_a/k_B T$, where $E_a$ is the activation energy, $k_B$ is the Boltzmann constant, and $T$ is the temperature. These RT behaviors keep similar variation tendency with increasing magnetic fields up to 14 T. Note that, the magnetic fields have been applied along three different crystallographic directions of the crystal, as mentioned in the *Experiments* section and schematically displayed in the upper inset of Fig. 2(a). With the magnetic fields applied in the $B_\perp$ configuration (i.e., *B* // [210]), the absolute value of the resistivity increases in the 1.8 – 300 K temperature region, showing a positive magnetoresistance ( MR = $\frac{\rho_B - \rho_0}{\rho_0}$ ) effect. Specifically, both $E_a$ and $T_{\min}$ increase with increasing magnetic fields. After fitting the resistivity data using the Arrhenius equation [Fig. 2(b)], $E_a$ at various magnetic fields were obtained and shown in the inset of Fig. 2(b). $E_a$ increases with the magnetic field monotonically for *B*≤5 T, and reaches a stable value of ~18 meV for *B*>5 T. The increase in $T_{\min}$ with the magnetic fields is consistent with the enhanced activation energy. Similar RT behaviors were observed in the same single crystal with the magnetic fields applied along the [100] direction (Supplementary Fig. S2).

Next, we measured the resistivity as function of the magnetic field at various fixed temperatures to study the MR effect from 1.8 to 300 K. Note that, MR is an even function of the magnetic field, which can be verified by the MR vs. *B* curves in the *T* ≥ 5 K region [Fig. 2(c) for *B* // [210] and 2(d) for *B* // [100]. We therefore only plotted the resistivity data in 0 – 14 T field region to analyze potential quantum oscillations in the 1.8 – 4 K temperature region. Following the dash arrows, one can see that the variation of the resistivity with the magnetic field is less pronounced with heating. The corresponding MR values calculated from the resistivity data were plotted in Supplementary Fig. S3 (*B* // [210]) and S4 (*B* // [100]), respectively. Interestingly, we observe oscillation patterns at low temperatures, similar to the previous observation in Te crystals, in which they observed the oscillation patterns at the same temperature region.[10] It is worth mentioning that, in their experiments, the crystals possess a quantum limit transition at ~ 4 T, above which, the log-periodic oscillations occurred. The log period is verified in *Ref.* 10, with applied magnetic field up to 53 T. The crystal growth condition, RT and MR behavior show high similarity, therefore, the log-periodic oscillations could also exist in the present Cr:Te crystals. Since the oscillation patterns follows sine or cosine function, a straightforward method to obtain oscillation is to take a secondary derivative of the resistivity, due to $cos''x = -cosx$. The negative secondary derivative resistivity, together with the resistivity versus temperature curves, at $T = 1.8$ K are shown in the insets of Fig. 2(c) and 2(d), respectively. As indicated by the blue arrows, the oscillations are much clear than those of the raw resistivity versus $B$ curves. The local maximum (peak) and minimum (valley) can be indexed following the aforementioned quantum oscillation models, e.g., the indices are linear $B$ for AB oscillations (the magnetic field that causes a phase difference of $2\pi$ should be the period, which is $\Delta B = h/eA$ for a ring of area $A$) with $1/B$ for the SdH oscillation, and linear with $logB$ for the logarithmic oscillation[8-10]. Therefore, the indexing rules are analyzed for the oscillation patterns at 1.8 K, as shown in Fig. 3. To simplify the fitting, integer and half integer numbers 1 – 2.5 are indexed to the peak/valley positions of oscillations for the magnetic field applied along the [100] and [210] directions, respectively. The indices are linearly fitted to $B$ [Fig. 3(a) and 3(d)], $logB$ [Fig. 3(b) and 3(e)], and $1/B$ [Fig. 3(c) and 3(f)], respectively. One can see that the linear fittings yield different merit factors using different quantum oscillation models. By comparing the adjusted $R$ square parameters [Fig. 3(g)] obtained from these fittings, one may evaluate the most appropriate models for the oscillations for $B$ // [210] and $B$ // [100], respectively. An interesting finding is that, the quantum oscillation along the [100] direction follows the logarithmic scale rule while the oscillation along the [210] direction follows the SdH rule.

Further, we plot the temperature dependent quantum oscillations with the best indexing rule of each for $B$ // [100] and $B$ // [210] in Fig. 4. We first focus on the oscillations for $B$ // [100] [Fig. 4(a) and 4(b)]. By conducting the secondary derivative, clear oscillation patterns with log$B$-periodicity are obtained, as shown in Fig. 4(b), where the $x$-axis has been set to logarithmic scale. This logarithmic quantum oscillation is similar to those observed in ZrTe$_5$[8], HfTe$_5$[9], and tellurium[10], and can be attributed to the manifestation of DSI, which indicates that the system has a geometric series of length scales. In detail, the hole band in these materials are Dirac-like fermions, which can be affected by the Coulomb potential generated by the electron charge impurities. The impurity induced long-range Coulomb potential possesses a continuous scaling symmetry.[8] Due to the quasi-bounded states between the relativistic fermions and the charge impurities, the energies of which constitute a geometric series, matching the general feature of DSI. The applied magnetic field enables the quasi-bounded states in different energy scales to shift to the Fermi level continuously. In this case, the logarithmic scale tuning of the magnetotransport properties is obtained due to the scattering between free carriers and bounded states. In practical, tellurium's Fermi level sits in the valence band and forms a dumbbell-like Fermi surface, with its longitudinal direction along the $c$ axis. The Te vacancies, on the other hand, can be treated as negative charge centers that provide Coulomb attractions on the positive relativistic fermions. For the logarithmic oscillations, the scale factor can be defined as $\lambda = B_{n+1}/B_n$, which is ~1.78 for the present Cr-doped tellurium single crystal, ~2.33 for tellurium single crystal,[10] and ~3.16 for ZrTe$_5$ single crystals.[8] The semiclassical quantization condition resulting in DSI is $\lambda = e^{2\pi/s_0}$, where $s_0 = \sqrt{\alpha^2 - 1}$ by assuming the charge number as 1, and $\alpha$ is the fine-structure constant.[8] For our crystals, the fine-structure constant $\alpha$ is ~2.04, much larger than 1/137 due to the small Fermi velocity. Using the equation $\alpha = e^2/4\pi\varepsilon_0\hbar v_F$, where $e$ is the charge of an electron, $\varepsilon_0$ is the dielectric constant in vacuum, and $\hbar$ is the reduced Planck constant, the Fermi velocity is estimated to be ~1.08×10$^{-6}$ m/s, which is slightly larger than that obtained by the DFT calculations.[19]

The observed SdH oscillations for $B$ // [210], as shown in Fig. 4(c) and 4(d) can be described by the Lifshitz-Kosevich (LK) formula, with the Berry phase being taken into account:

$$\frac{\Delta\rho}{\rho(0)} = \frac{5}{2}\left(\frac{B}{2F}\right)^{\frac{1}{2}} R_T R_D R_S \cos\left[2\pi\left(\frac{F}{B} + \gamma - \delta\right)\right]$$

Where $R_T = \xi T\nu/B\sinh(\alpha T\mu/B)$, $R_D = \exp(-\xi T_D\nu/B)$, and $R_S = \cos(\xi g\nu/2)$. Here, $\nu = m^*/m_e$ is the ratio of the effective cyclotron mass $m^*$ to the free electron mass $m_e$; $g$ is the g-factor; $T_D$ is the Dingle temperature; and $\xi = (2\pi^2 k_B m_e)/\hbar e$, where $k_B$ is Boltzmann constant, $\hbar$ is the reduced Planck constant, and $e$ is the charge of an electron. The oscillation of $\Delta\rho$ is described by the cosine term with a phase factor $\gamma - \delta$, in which $\delta = 0$ for 2D Fermi pockets, and $\pm 1/8$ for 3D Fermi pockets, $\gamma = 1/2 - \Phi_B/2\pi$, where $\Phi_B$ is the Berry phase. The Berry phase of the SdH oscillations can also be obtained via the Landau fan diagram, which is the extrapolated Landau-level index $n$ at the extreme field limit $1/B \to 0$.

Since $\rho_{xx} \gg \rho_{xy}$ in our measurements, we assign the maximum and minimum of oscillations as integer Landau indices and half integer Landau indices, respectively, and carefully fit the data linearly.[23,24] As shown in Supplementary Fig. S5, the y-axis-intercept of the linear fitting is ~0.65±0.05 (the unit is 2π), which indicates the pi Berry phase. The slop of the fitting is 12.1 T, which is the frequency (*F*) of the oscillations. According to the Onsager-Lifshitz equation $F = (\varphi_0/2\pi^2)A_F$, where $A_F$ is the extremal area of the cross-section of the Fermi surface perpendicular to the direction of the magnetic field, and $\varphi_0$ is the magnetic flux quantum, the Fermi pocket area is ~0.0012 Å$^{-2}$. According to the LK formula, the effective mass of carriers contributing to the SdH oscillation can be obtained through fitting the temperature dependent oscillation amplitude to the thermal damping factor $R_T$, which is shown in Supplementary Fig. S6. During the effective mass fitting, we employ the intensity of the lowest Landau level peak ~ 12.1 T, which yields an effective mass $m^* \sim 1.50 \pm 0.08\ m_e$. Thus, one can obtain the Fermi velocity $v_F = \hbar k_F/m^* \sim 2.5 \times 10^5$ m/s for the Cr-doped tellurium single crystal. Moreover, we can estimate the effective mass from the log-periodic oscillations using the $R_T$ relationship in SdH oscillations, as shown in Supplementary Fig. S6, which is ~ $1.65 \pm 0.05\ m_e$. A comparison of the fitting parameters obtained from the present Cr:Te and Te crystals reported in Ref. 10 and 12 are shown in Table 1.

Table 1: The parameters derived from quantum oscillations.

|  | $F$ (T) | $A$ (Å$^{-2}$) | $k_F$ (Å$^{-1}$) | $v_F$ (m/s) | $m^*$ ($m_e$) |
|---|---|---|---|---|---|
| *This work* | 12.1 | 0.0012 | 0.02 | 2.5×10$^5$ | 1.50±0.08 |
| Bridgeman Te crystal[12] | 4.6 |  |  |  | 0.06 |
|  | 68 |  |  |  | 0.3 |
| PVD Te crystal[10] | 4.4 |  |  | 2.9×10$^5$ |  |
|  |  |  |  | 1.9×10$^5$ (DFT) |  |

Further, we measured the quantum oscillations at $T = 1.8$ K at various fixed rotation angles. During the measurements, the direction of the magnetic field rotates from the [100] direction to the [210] direction, within the *ab* plane of the crystal. At each angle, the resistivity was measured by sweeping the magnetic field from 0 to 14 T. Note that, the resistivity for different angles were shown separately by adding some constants to each curve, because the resistivity versus *B* curves are nearly overlapped with each other (Supplementary Fig. S7). Regardless of the rotation angle, the resistivity is exactly the same in the low field region ($B < 4.5$ T) (Supplementary Fig. S7), which indicates weak anisotropy in MR within the *ab* plane of the crystal. However, for $B > 4.5$ T, the quantum oscillations are strong enough to show difference for different angles, as shown separately in Fig. 5(a). The oscillation patterns in the form of secondary derivative resistivity vs. *B* curves are plotted in Fig. 5(b), where constants are also added to each curve to separate them for clarity. One important feature of these oscillations is the quasi-60° rotation symmetry, i.e., the 0° and 60° oscillation curves are similar, as shown in Supplementary Fig. S8. This quasi-60° rotation symmetry can be understood by the crystallographic symmetry. In a trigonal lattice, the [100], [010] and [$\bar{1}\bar{1}0$] are equivalent, for the rotation angle $\theta=60º$, the direction of the magnetic field is along the [110] direction, which is the opposite direction of [$\bar{1}\bar{1}0$] [Fig. 1(b)].

Moreover, we conducted the rotation-angle-dependent MR measurements within the *ac* or (*h0l*) plane, which means that the direction of the magnetic field rotates from the [001] direction to the [100] direction [see the inset in Fig. 6(a)]. Zhang *et al.* [10] reported that tellurium single crystals show negative magnetoresistance (maximum value ~ -22% at 14 T and 25 K) when the direction of the magnetic field is parallel to the electric current (i.e., *B* // [001]), which is attributed to the chiral anomaly in Weyl fermion[25,26] system. As shown in Ref.

9, Te single crystals show chiral anomaly dominant negative longitudinal MR, which can be described by $\sigma(B) = (1 + C_w B^2) \cdot \sigma_{WAL} + \sigma_N$, where $C_W$ is a positive parameter representing the chiral anomaly contribution to the conductivity, the $\sigma_{WAL}$ and $\sigma_N$ are the conductivity imparted by the weak antilocalization effect and conventional nonlinear band contributions around the Fermi level, respectively. As shown in Supplementary Fig. S9, the magnetoconductance can be fitted with the equation fairly well. For our Cr:Te single crystals, similar negative MR is observed for *B* // [001] // *I*, as shown in Fig. 6 and Supplementary Fig. S10. We noticed that the anomalous temperature dependent relationship has been observed in both Ref. 10 (below 25 K) and here (below 20 K). Therefore, we focus on the low-temperature angular-dependent longitudinal MR in this section. Note that there are ten MR vs. *B* curves at each temperature in Fig. 6 and the angle gradient is 10°. The negative MR appears for *T*≤50 K [Supplementary Fig. S10] and is most pronounced for *T* = 20 K [Fig. 6(d)], and it increases with increasing *B* from 0 T and exhibits a local maximum value at a certain *B*, then decreases with further increase in *B*, e.g., the maximum negative MR is ~ -6% at *T* = 50 K and *B* = 11.6 T, ~ -19% at *T* = 30 K and *B* = 9.4 T, ~ -27% at *T* = 20 K and *B* = 8 T. Further, the negative MR effect changes significantly as the direction of the magnetic field rotates from the [001] direction to the [100] one (Fig. 6), which is particularly significant for *T* = 20 and 25 K [Fig. 6(d) and 6(e)], e.g., the negative MR gradually disappears with increasing angle from 0 to 90º; the values of the magnetic fields where the negative maximum MR occurs gradually decreases with increasing angle from 0 to 90º.

As shown in Fig. 2(a) and Supplementary Fig. S2, there seems no signs of negative MR effect whether *B* // [210] or [100] direction, one may still argue that the ~7.7% Cr dopants in the tellurium single crystal could introduce ferromagnetism and negative MR due to reduced magnetic scatterings. We therefore conducted some measurements in low magnetic field region and at low temperatures to search for the possible ferromagnetism and negative MR. As shown in Fig. 7(a), for *B* // [100] direction, the MR vs. *B* curve shows a butterfly shape at 1.8 K, which couples the magnetic hysteresis (MH) loop at 1.8 K (diamagnetic background has been subtracted). The MH loop is saturated at ~ 0.15 T, where the negative MR reaches the maximum value. More low-temperature magnetic hysteresis (MH) loops are shown in Fig. S11. The appearance of negative MR suggests magnetic scattering's dominance in such a low field region. For larger magnetic field (*B* > 2 T), we observed the positive MR and quantum oscillations, as shown in the inset of Figs. 2(c) and 2(d), and Supplementary Fig. S3 and S4. The butterfly shape resistivity vs. *B* curve is also observed at higher temperatures up to 3.4 K, however, the relative resistivity changes decrease with increasing temperature, as shown in Fig. 7(b). At *T* = 3.8 K, the negative MR is absent, suggesting that the ferromagnetism coupled with the MR behavior vanishes at *T*=3.8 K. The ferromagnetism in such low temperatures (<3.8 K) and a low field region (<0.15 T) should not be the reason for the negative MR displayed in Fig. 6 where the negative MR originates from the chiral anomaly in materials system hosting Weyl fermions.

We further studied the carrier's properties based on the Hall effect in the 1.8 to 300 K temperature region, as shown in Fig. 8. The Hall resistivity vs. *B* curves are linear with the magnetic fields, with a positive slope, suggesting the hole carrier's dominance of the electronic transport properties. We calculate the Hall coefficient $R_H$ and obtain the hole carrier density via $n_H = 1/R_H e$ (*e* is the charge of an electron) and Hall mobility $\mu_H = R_H \sigma$ ($\sigma$ is the conductivity). The Hall mobility and carrier density are plotted against temperature in Fig. 8(b). The carrier density is nearly constant below 200 K, with a value of ~ $4\times10^{16}$ cm$^{-3}$. The mobility, however, changes with temperature dramatically, e.g., ~230 cm$^2$/Vs at 1.8 K, and increases to ~1320 cm$^2$/Vs at 30 K, and further deceases to ~ 350 cm$^2$/Vs at 300 K. It is noted that, in low temperature region, the carrier density keeps almost constant, regardless of the significant thermal activation behaviors of the RT curves.

## Conclusions

Transition metal element of Cr was successfully doped into narrow-bandgap semiconductor of tellurium single crystals via the physical vapor transport method. The 7.7 % Cr-doped Te crystals show ferromagnetism below 3.8 K temperature region, where coupled magnetic hysteresis loop and butterfly-shaped magnetoresistance *vs*. magnetic field curves are observed, implying intrinsic ferromagnetic ordering in the Cr:Te crystals. With successful magnetic doping, the single crystals still present similar transport property with pristine tellurium, e.g., the quantum limit, and quantum oscillations in MR curves. Interestingly, the quantum oscillations show crystallographic orientation dependence, e.g., the oscillations following logarithmic oscillations for the direction of the magnetic field *B* parallel to the [100] direction (*B* // [100]), and SdH oscillations for *B* // [210] direction. The low-temperature ferromagnetism ($T \leq 3.8$ K) is not responsible for the relatively larger temperature- and rotation-angle-dependent negative MR effect at higher temperatures (e.g., 27% at *T* = 20 K and *B* = 8 T) when the direction of *B* is along the electric current direction. Our results demonstrate a successful ferromagnetic doping in elemental semiconductor tellurium, which may inspire more magnetic tellurium based electronic studies.

## Acknowledgements

This work is supported by National Natural Science Foundation of China (Grant No. 11974155) and Doctor Foundation of Jiangxi Science and Technology Normal University (Grant No. 2022BSQD34). WZ and JK acknowledge the supporting form ARC Centre of Excellence in Future Low-Energy Electronics Technologies No. CE170100039 and the Australian Research Council Discovery Project DP200102477.

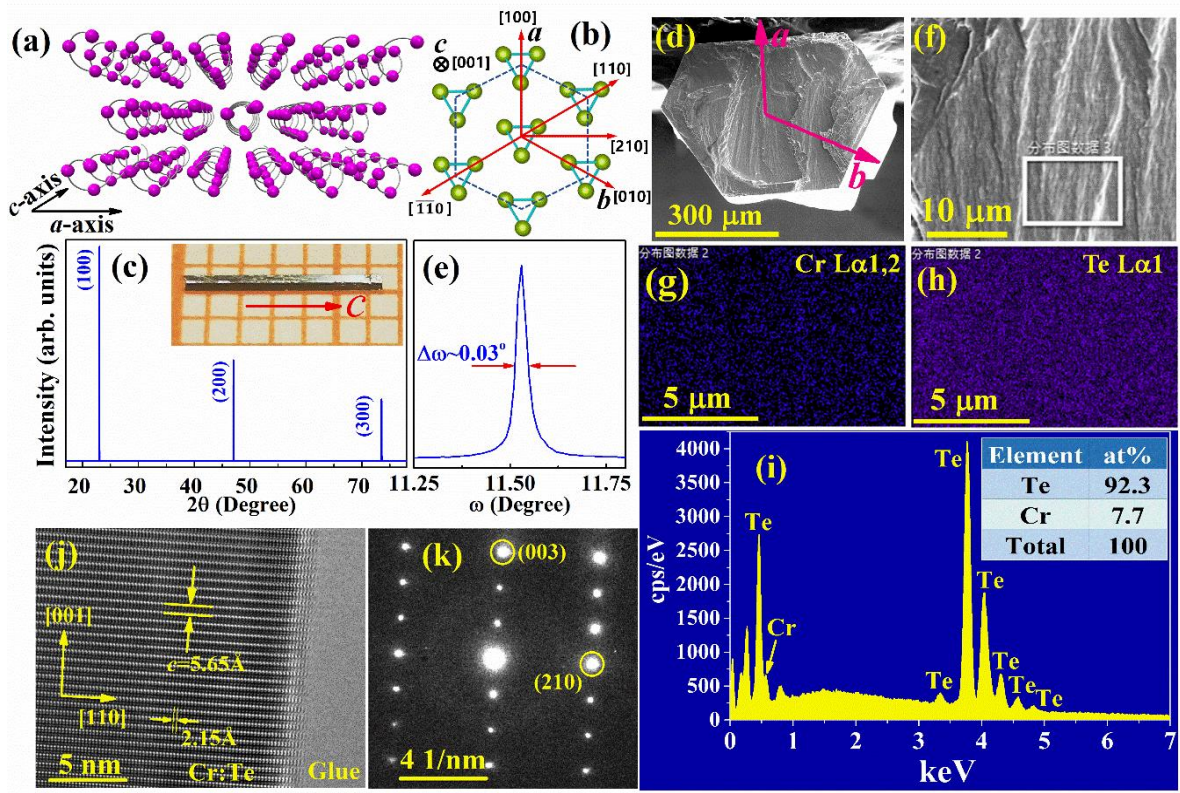

**Fig. 1.** Crystal structure of Cr:Te single crystal. (a,b) A sketch of the crystal structure of tellurium. (c) XRD pattern taken on a shiny surface. (d) a cross sectional SEM image. (e) XRD rocking curve taken on the (100) diffraction peak. (f) a cross sectional SEM zoom-in image, (g,h) EDS elemental mapping of the Cr and Te elements, which was taken on a relatively large area of the white box shown in (f). (i) Element spectrum, (j) HRTEM image, (k) SAED pattern.

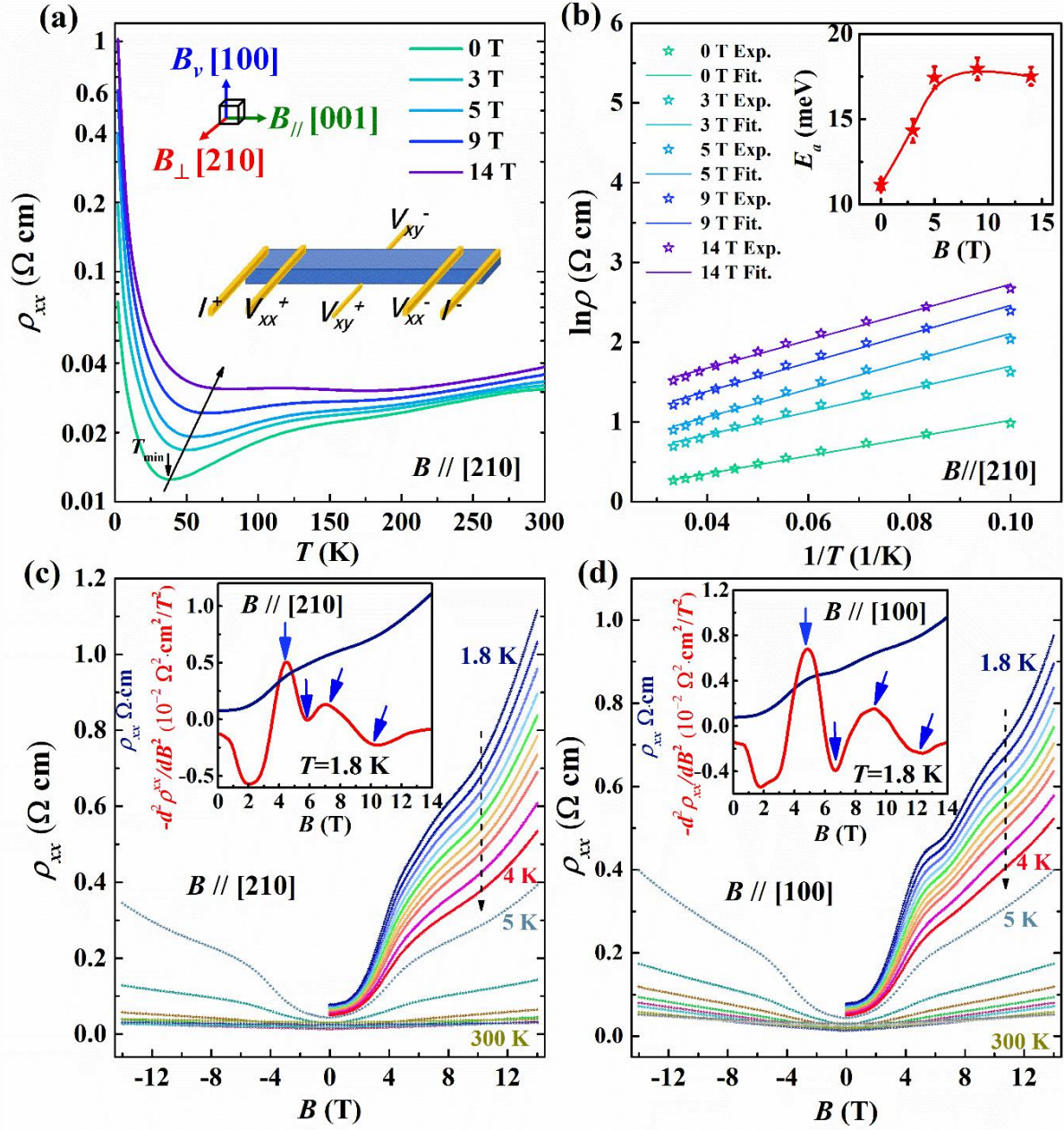

**Fig. 2.** Magnetotransport properties of the Cr:Te single crystal. (a) Temperature dependence of the resistivity in magnetic fields up to 14 T, for $B \; // \; [210]$. (b) The fitting of the low-temperature resistivity for $B \; // \; [210]$ using the Arrhenius equation. The obtained activation energy is plotted in the inset. (c) and (d) Resistivity as a function of the magnetic field for $B \; // \; [210]$ and $B \; // \; [100]$, respectively. The insets show the temperature dependence of the resistivity in zero magnetic field and corresponding second derivative of the resistivity showing quantum oscillations at 1.8 K.

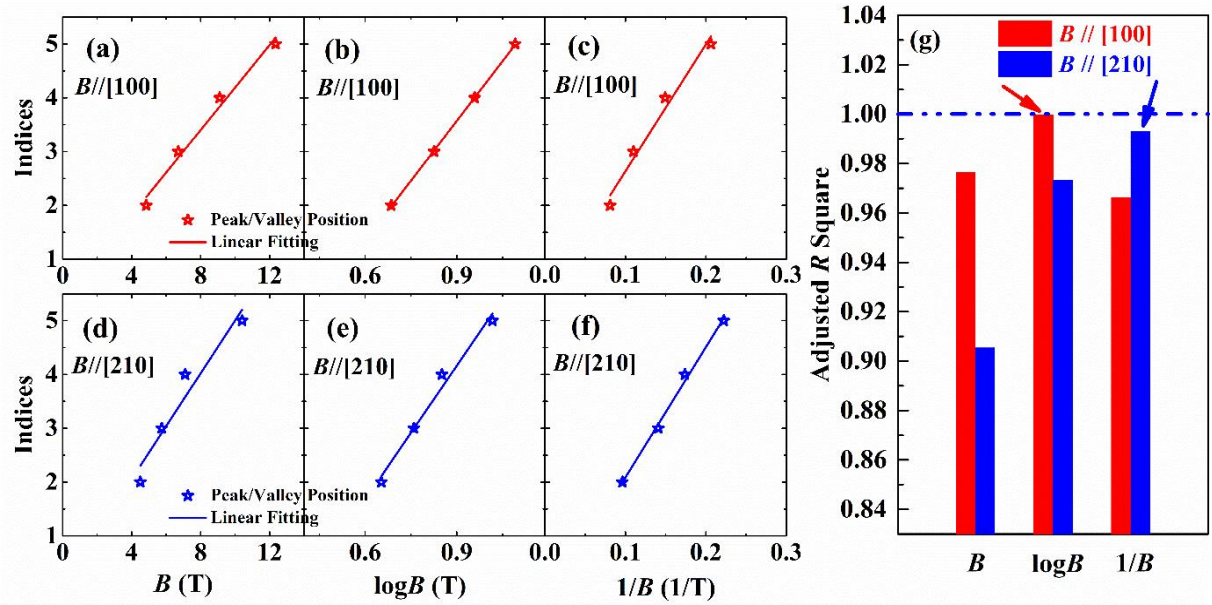

**Fig. 3.** The fitting of the indices obtained from the quantum oscillations observed at 1.8 K for $B$ // [100] (a-c) and $B$ // [210] (d-f), based on different quantum oscillation models. (g) The adjusted $R$ square values for different fitting models.

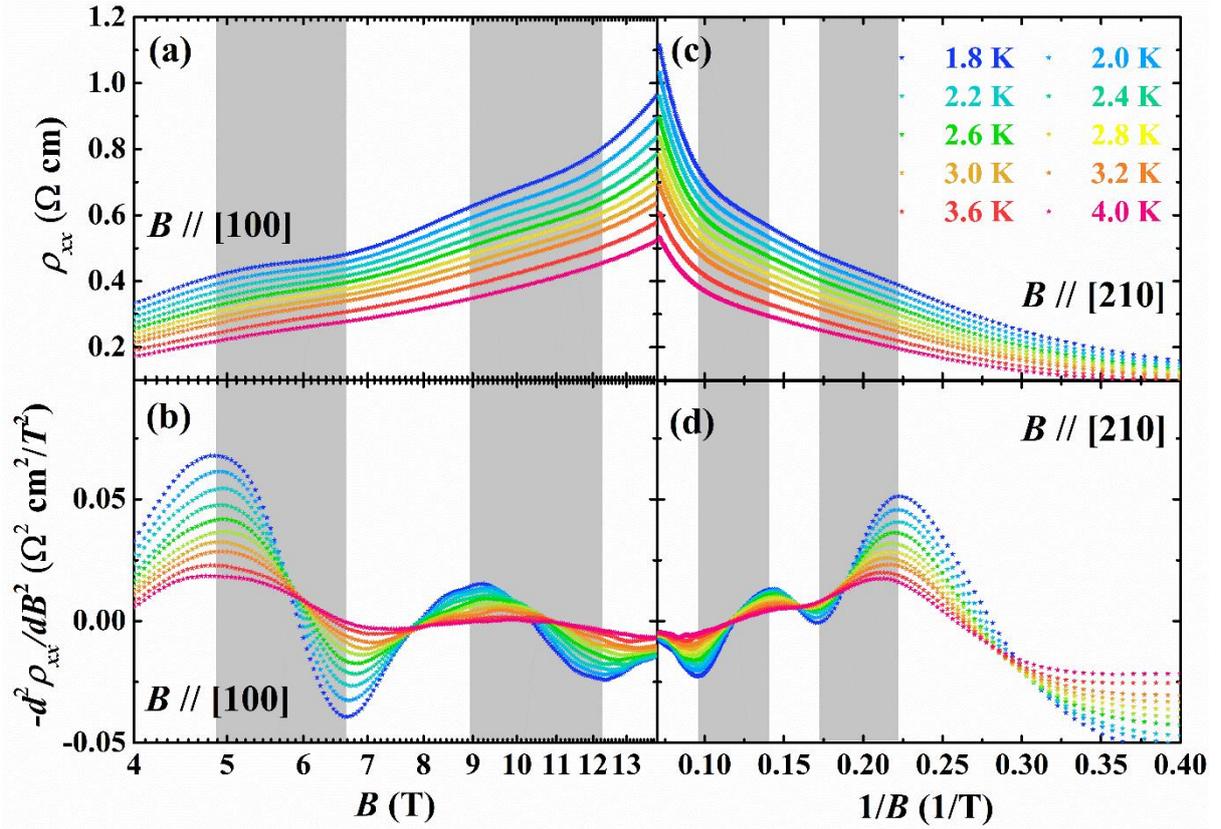

**Fig. 4.** Temperature dependent quantum oscillation patterns. (a,b) The resistivity and secondary derivative resistivity as a function of the magnetic field applied along the [100] direction. (c,d) The resistivity and secondary derivative resistivity as a function of the magnetic field applied along the [210] direction.

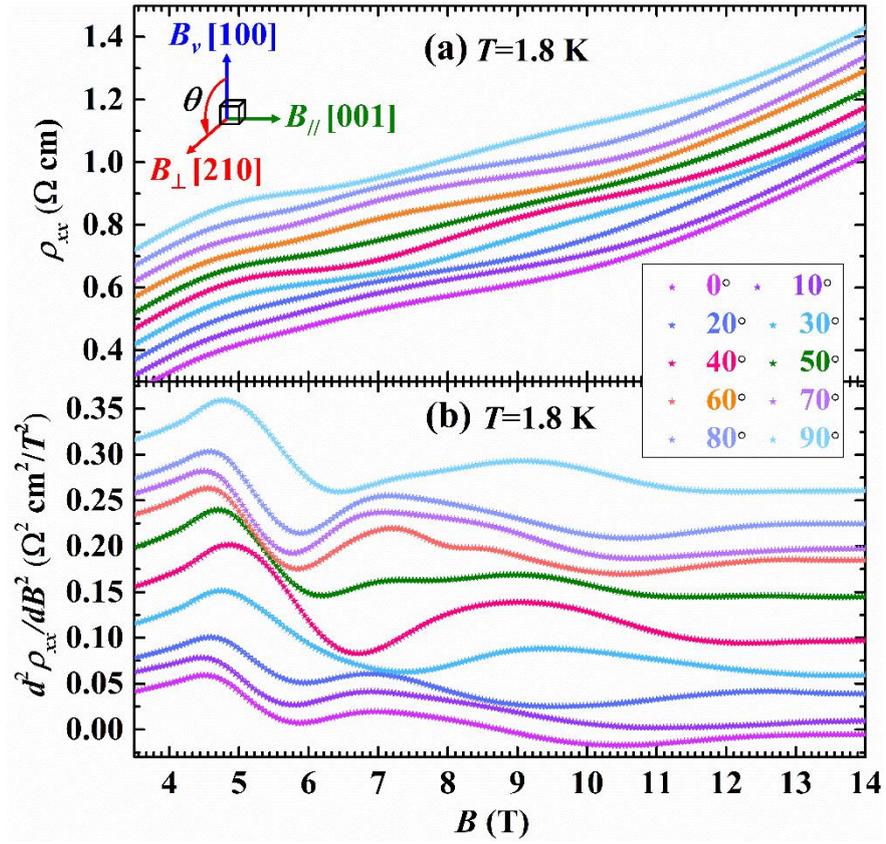

**Fig. 5.** The quantum oscillations are plotted at different rotation angles, where the resistivity and the secondary derivative resistivity are shown in panel (a) and (b), respectively.

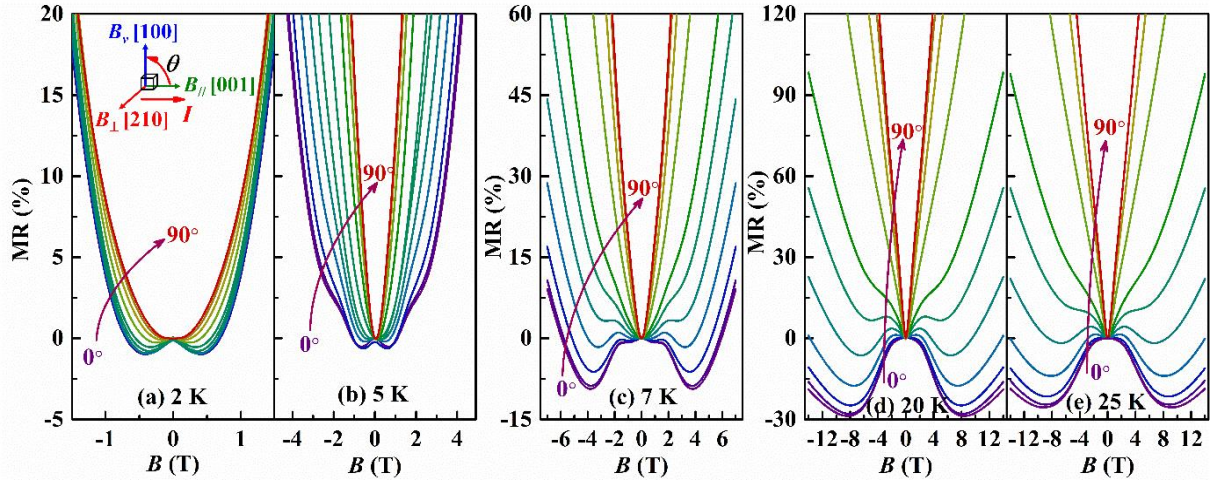

**Fig. 6.** The rotation angle dependent *MR* vs. *B* curves at fixed temperatures of 2 K (a), 5 K (b), 7 K (c), 20 K (d), and 25 K (e). The direction of the magnetic field rotates from [001] to [100] within the (*h*0*l*) plane. The MR values have been divided by 3 at 7 K, and divided by 5 at 20 and 25 K, to adapt the MR dimension at low temperatures.

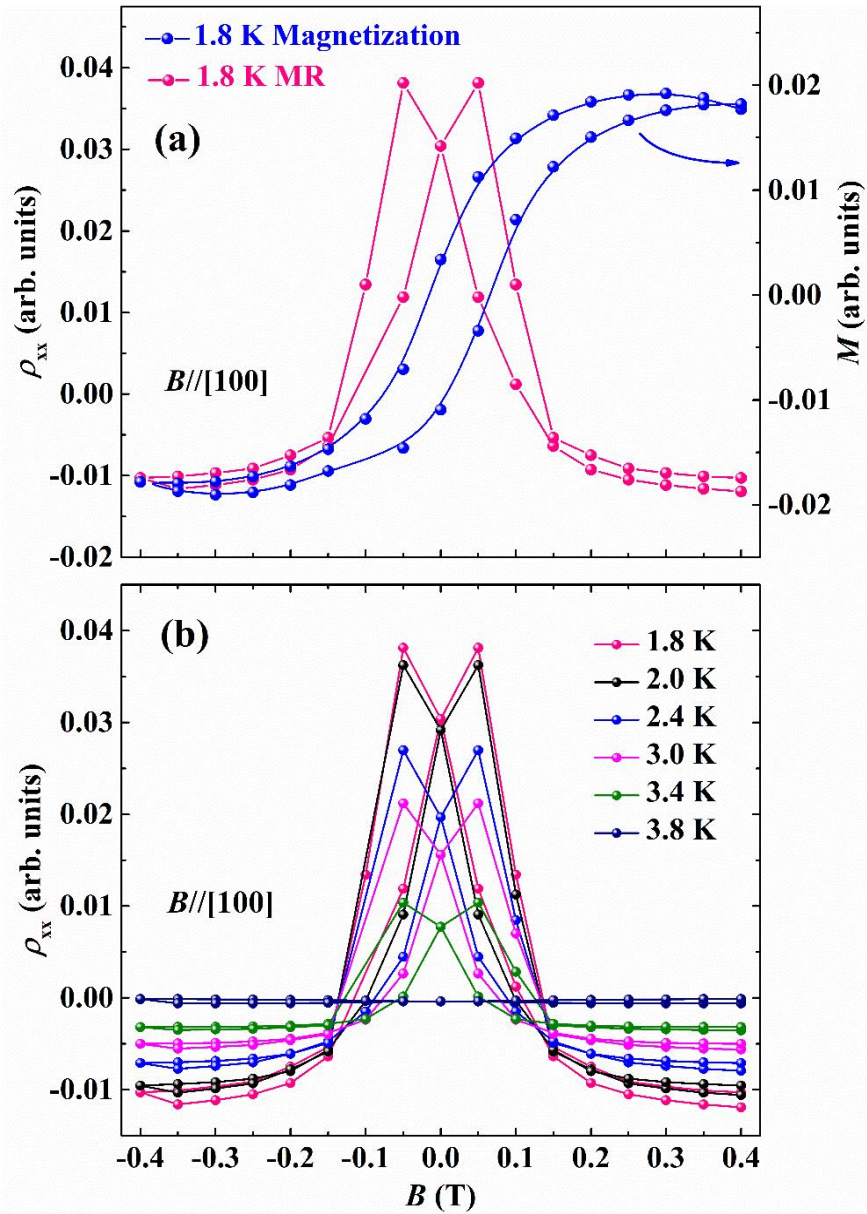

**Fig. 7.** (a) The butterfly MR effect at 1.8 K and magnetic hysteresis loop in the low field region, (b) The butterfly MR effect at 1.8 K at various temperatures.

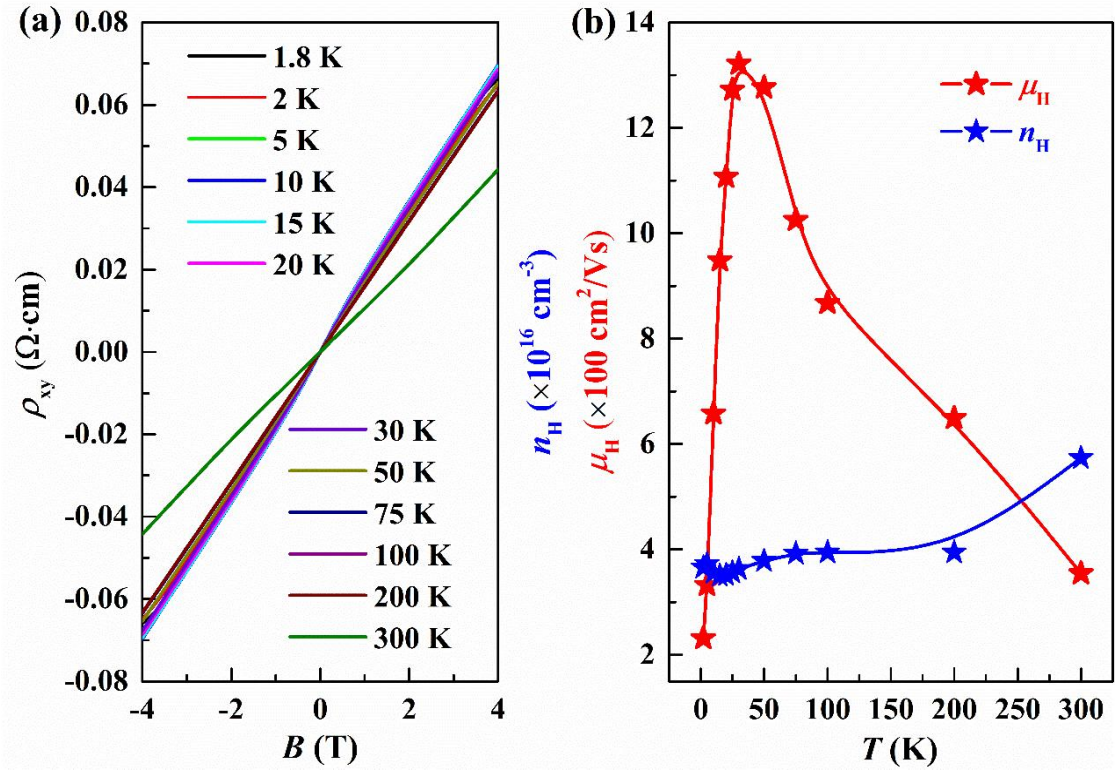

**Fig. 8.** The Hall effect in the 1.8 – 300 K region. (a) The Hall resistivity vs. $B$ curves at various temperatures. (b) The carrier's density and mobility are plotted against temperature from 1.8 to 300 K.